\documentstyle[twocolumn,epsf]{jpsj}
\def\runtitle{Possible Spin-Singlet Superconductivity in (TMTSF)$_2$X}
\def\runauthor{Mitake {\sc Miyazaki}, Keita {\sc Kishigi} and
 Yasumasa {\sc Hasegawa}}
\title{Possible Spin-Singlet Superconductivity in (TMTSF)$_2$X: 
Superconducting Transition Temperature in a Magnetic Field}
\author{Mitake {\sc Miyazaki},  Keita {\sc Kishigi} and
 Yasumasa {\sc Hasegawa}}
\inst{Faculty of Science, Himeji Institute of Technology,
\\Kamigouri-cho, Akou-gun, Hyogo 678-1297}
\recdate{\today}
\abst{ 
 We study the transition temperature $T_{\rm c}(H)$ of a
quasi-one-dimensional (Q1D) superconductivity 
which is derived from the quantum effect of an electron motion 
in a strong magnetic field. We calculate $T_{\rm c}(H)$ of both
isotropic and anisotropic superconductivity by taking account of the
optimal momentum of the Cooper pairs and the 
effect of higher harmonic terms along second conducting axis in the
tight-binding model. We find that, 
although $T_{\rm c}(H)$ of the spin-singlet superconductivity 
is suppressed strongly 
by the Zeeman effect, the suppression of $T_{\rm c}(H)$ is not very severe
if we take the optimal pair-momentum and the   
higher harmonic terms into account. The obtained $T_{\rm c}(H)$ for the
spin-singlet superconductivity  
is consistent with the experimental 
results in TMTSF salts. 
}
\kword{organic superconductors, 
quasi-one-dimension, Pauli paramagnetic limit}
\begin{document}
\sloppy
\maketitle

It is known that the spin-singlet superconductivity is
destroyed by both the orbital frustration and the Pauli paramagnetic
effect. Recently, 
the reentrance of the superconductivity which is caused by quantum 
effect of orbital motions along the open Fermi surfaces in a strong 
magnetic field has attracted 
interest~\cite{rf:1,rf:2,rf:3,rf:4,rf:5,rf:6,rf:7}. 
The anomaly of the resistivity in a
strong magnetic field has been observed~\cite{rf:8,rf:9,rf:10} in 
quasi-one-dimensional (Q1D) organic superconductors (TMTSF)$_2$X
(where anion X is ClO$_4$ or PF$_6$) when the magnetic field is applied
along the second conducting axis ($b$ axis).
This is thought to be a signal of the superconductivity in a strong
magnetic field, since the critical 
temperature $T_{\rm c}(H)$ exceeds both the upper critical field 
$H_{\rm c2}$ derived in GL theory~\cite{rf:11,rf:12} and Pauli 
paramagnetic limit $H_{\rm P}$[T]$=1.84T_{\rm c}$[K]~\cite{rf:13,rf:14}.  

In the Q1D system,   
it has been known~\cite{rf:1,rf:2,rf:3,rf:4,rf:5,rf:6} that the  
spin-singlet superconductivity 
is not destroyed completely due to the Zeeman effect by constructing
the Larkin-Ovchinnikov-Fulde-Ferrell (LOFF) superconducting 
state~\cite{rf:15,rf:16} 
in which Cooper pair is formed by the electrons
$({\mib k},\uparrow)$ and 
$(-{\mib k}+{\mib q},\downarrow)$.
The electron 
$(-{\mib k}+{\mib q},\downarrow)$ can be on the down-spin Fermi 
surface for any  
$({\mib k},\uparrow)$ on the up-spin Fermi surface in a 1D system by
choosing the appropriate ${\mib q}$, which is similar to the nesting 
of the Fermi surface
in the spin-density-wave (SDW) case. Even in the Q1D system, the ``nesting"
condition for the LOFF state was thought to become perfect in the strong
magnetic field~\cite{rf:1,rf:2,rf:3,rf:4,rf:5,rf:6,rf:7}. These theoretical
calculations~\cite{rf:1,rf:2,rf:3,rf:4,rf:5,rf:6,rf:7} for $T_{\rm c}(H)$ 
based on the approximation that the Fermi velocity is independent of the
position of the Fermi surface.

Recently, Lebed~\cite{rf:17} has shown that the 
nonlinearity effect of the energy dispersion along the $a$ axis on the
Zeeman splitting causes the finite upper critical field in Q1D systems. 
He obtained that the critical magnetic field for the LOFF state is 
$H^{\rm LOFF}_{\rm P}\sim  0.6\sqrt{t_a/t_b}H_{\rm P}$, where $t_a$ and 
$t_b$ are the hopping matrix elements along $a$ and $b$ axes,
respectively. Applying this result to (TMTSF)$_2$X, $H^{\rm LOFF}_{\rm
P}\sim 4$Tesla is obtained, which is smaller than the experimentally
observed value by Lee {\it et al}. (at least $7$T)~\cite{rf:8,rf:9}.
This result may suggest that the superconductivity in this system is a 
spin-triplet state which is not affected by the Zeeman effect. 
However, $T_{\rm c}(H)$ measured by Lee {\it et al}. does not reveal the 
reentrant behavior expected in the case of spin-triplet 
superconductivity.~\cite{rf:1,rf:2,rf:3,rf:4,rf:5} 
 
In this paper, we study $T_{\rm c}(H)$ of a Q1D spin-singlet superconductor
by taking the optimal pair momentum ${\mib q}$ and introducing the 
higher harmonic terms along the  
second conducting axis in the tight-binding model, 
which has been discussed in the field-induced-spin-density-wave
(FISDW)~\cite{rf:18,rf:19,rf:20} and the quantum hole effect 
(QHE)~\cite{rf:21,rf:22} in FISDW. We consider both isotropic and
anisotropic pairing states for spin-singlet superconductivity, since the
symmetry of the pairing is still controversial~\cite{rf:23,rf:24}.

We consider the anisotropic tight-binding model including the   
effect of the Zeeman
splitting (we take $\hbar=k_{\rm B}=c_0=1$, where
$c_0$ is the velocity of light):
\begin{eqnarray}
E_{{\mib k},\sigma}&=&-2t_a\cos(ak_x)-2t_b\cos(bk_y)
-2t_{2b}\cos(2bk_y)\nonumber\\
& &-2t_{3b}\cos(3bk_y)-2t_{4b}\cos(4bk_y)-2t_c\cos(ck_z)\nonumber\\
& &-\sigma\mu_{\rm B}H-\mu,
\end{eqnarray}
where 
$t_a\gg t_b\gg |t_{2b}|>|t_{3b}|>|t_{4b}|\sim t_c$, $\sigma\mu_{\rm B}H$
is the Zeeman energy for $\uparrow(\downarrow)$ spin $(\sigma=+(-))$ and
$\mu$ is the chemical potential to give the quarter filled electrons.

The orbital effect of the magnetic field is treated 
by the Peierls substitution, i.e. 
${\cal H}_0=E({\mib k}\rightarrow -i\nabla-e{\mib A})$. We take the
vector potential ${\mib A}$ as ${\mib A}=(0,0,-Hx)$. Since $t_a\gg t_b$,
we can linearize the 
energy dispersion along $k_x$ for each $k_y$. Then the eigenvalues 
are given by
\begin{equation}
\epsilon^{\alpha}_{k_x,k_y,\sigma}=v^{\rm F}_{k_y,\sigma}(\alpha k_x
-k^{\rm F}_{k_y,\sigma}),
\end{equation}
where $\alpha={\rm sgn}(k_x)$ refers to the right/left sheet of the 
Fermi surface. Although we linearize the dispersion, we consider the
$k_y$ and $\sigma$ dependence of the Fermi wave number 
$k^{\rm F}_{k_y,\sigma}$ and the Fermi velocity along the $a$ axis,
$v^{\rm F}_{k_y,\sigma}=2t_aa\sin(ak^{\rm F}_{k_y,\sigma})$. Note that
$k^{\rm F}_{k_y,\sigma}$ and $v^{\rm F}_{k_y,\sigma}$ do not 
depend on $k_z$ when the
magnetic field is applied along the $b$ axis. For given $k_y$
and $\sigma$, $k^{\rm F}_{k_y,\sigma}$ is obtained by,
\begin{eqnarray}
k^{\rm F}_{k_y,\sigma}&=&\frac{1}{a}\cos^{-1}\left[(2t_b\cos (bk_y)
+2t_{2b}\cos (2bk_y)\right.\nonumber\\
&+&2t_{3b}\cos (3bk_y)+2t_{4b}\cos (4bk_y)\nonumber\\
&+&\left.\sigma\mu_{\rm
B}H+\mu)/t_a\right].
\end{eqnarray}
The corresponding eigenstates are 
\begin{equation}
\phi^{\alpha}_{{\mib k},\sigma}({\mib r})={\rm e}^{{\rm i}{\mib k}\cdot 
{\mib r}}
\sum_n {\rm e}^{{\rm i}n(ck_z-Gx)}J_n(\eta^{\alpha}_{k_y,\sigma}),
\end{equation}
where $G=eHc$, $J_n(\eta^{\alpha}_{k_y,\sigma})$ is Bessel function and 
$\eta^{\alpha}_{k_y,\sigma}
=-2\alpha t_c/v^{\rm F}_{k_y,\sigma}G$. 

If we expand $E_{{\mib k},\sigma}$ to the second order in  
$t_b/t_a$ around $k_x\sim \pm k_{\rm F}$ in eq. (1), we obtain
\begin{eqnarray}
E_{k_y,\sigma,\alpha}&\approx&v_{\rm F}(\alpha k_x-k_{\rm F})
-2t_b\cos(bk_y)-2t_{2b}\cos(2bk_y)\nonumber\\
&-&2t_{3b}\cos(3bk_y)-2t_{4b}\cos(4bk_y)-\beta t_b\cos^2(bk_y)\nonumber\\
&+&\sigma\mu_{\rm B}H(1-\beta\cos(bk_y))-2t_c\cos(ck_z)-\mu,
\end{eqnarray}
where $k_{\rm F}$ and $v_{\rm F}=2t_a\sin(ak_{\rm F})$ are the Fermi 
velocity and the Fermi
wave number in the 1D case ($t_b=0$), respectively and
$\beta=\sqrt{2}t_b/t_a$. Lebed has discussed that the term proportional
to $\beta$ in the Zeeman energy causes the finite critical field in Q1D
systems. In the following, we do not use the expansion of 
$E_{{\mib k},\sigma}$ around $\pm k_{\rm F}$ (eq. (5)). We calculate 
$T_{\rm c}(H)$ by using eqs. (2)$\sim$(4).

The one-particle Green's function in the mixed representation is 
\begin{eqnarray}
& &G^{\alpha}_{\sigma}(x,x',k_y,k_z;\omega_n)=\sum_{n,n'}
{\rm e}^{{\rm i}n(ck_z-Gx)+{\rm i}n'(ck_z-Gx')}\nonumber\\
& &\times
J_n(\eta^{\bar{\alpha}}_{k_y,\sigma})
J_{n'}(\eta^{\alpha}_{k_y,\sigma})
\sum_{k_x}
{\rm e}^{{\rm i}k_x(x-x')}\tilde{G}^{\alpha}_{\sigma}(k_x,k_y;\omega_n),
\end{eqnarray}
where $\tilde{G}^{\alpha}_{\sigma}(k_x,k_y,\omega_n)=1/({\rm i}
\omega_n
-\epsilon^{\alpha}_{k_x,k_y,\sigma})$ and $\omega_n=(2n+1)\pi T$ 
is a Matsubara frequency.

We first study $T_{\rm c}(H)$ of the isotropic superconductivity caused by the
on-site attractive interaction $\lambda$  
in the mean field approximation. The linearized gap equation for 
an isotropic superconductivity is obtained as~\cite{rf:2,rf:3,rf:5}
\begin{eqnarray}
& &\Delta(x)=\lambda T\sum_{k_y,k_z}\sum_{\alpha,\omega_n}{\int}_{|x-x'|>d}
{\rm d}x'\Delta(x')\nonumber\\
& &\times G^{\alpha}_{\sigma}(x,x',k_y,k_z;\omega_n)G^{\bar{\alpha}}
_{\bar{\sigma}}(x,x',-k_y,-k_z;-\omega_n),
\end{eqnarray} 
where $d$ is the cutoff. 

The solutions of the gap equation (7) are written as  
\begin{equation}
\Delta_Q(x)={\rm e}^{{\rm i}Qx}\sum_l\Delta^Q_{2l}
{\rm e}^{{\rm i}2lGx},
\end{equation}
where Bloch wave vector $Q$ is taken as $-G<Q\leq G$. Then eq. (7) 
is written as a matrix equation
\begin{equation}
\Delta^Q_{2l}=\lambda\sum_{l'}\Pi^Q_{2l,2l'}\Delta^Q_{2l'},
\end{equation}
where 
\begin{equation}
\Pi^Q_{2l,2l'}=\sum_N\sum_{k_y}S^N_{l,l'}(k_y)\tilde{K}_{k_y}(Q+NG).
\end{equation}
The coefficients $S^N_{l,l'}(k_y)$ for an isotropic 
superconductivity are defined by
\begin{eqnarray}
S^N_{l,l'}(k_y)&=&\sum_nJ_{n+l}
(\eta^{\alpha}_{k_y,\sigma})
J_{n+l'}(\eta^{\alpha}_{k_y,\sigma})\nonumber\\
&\times&J_{n-l+N}(\eta^{\alpha}_{k_y,\bar{\sigma}})
J_{n-l'+N}(\eta^{\alpha}_{k_y,\bar{\sigma}}).
\end{eqnarray}
In the above $\tilde{K}_{k_y}(q_x)$ is given by
\begin{eqnarray}
\tilde{K}_{k_y}(q_x)&=&T\sum_{\omega_n}\sum_{\alpha,k_x}
\tilde{G}^{\alpha}_{\sigma}(k_x,k_y;\omega_n)
\tilde{G}^{\bar{\alpha}}_{\bar{\sigma}}(q_x-k_x,-k_y;-\omega_n)
\nonumber\\
&=&\sum_{\alpha}\frac{2}{v^{\rm F}_{k_y,\sigma}
+v^{\rm F}_{k_y,\bar{\sigma}}}\left[{\rm ln}
\left(\frac{2\Omega\gamma}
{\pi T}\right)+\Psi\left(\frac{1}{2}\right)\right.
\nonumber\\
&-&\left.Re\Psi\left(\frac{1}{2}+\frac{V^{\rm F}_{k_y,\sigma}}{4{\rm i}\pi T}
(k^{\rm F}_{k_y,\sigma}-k^{\rm F}_{k_y,\bar{\sigma}}+\alpha q_x)
\right)\right],
\end{eqnarray}
where $V^{\rm F}_{k_y,\sigma}=
2v^{\rm F}_{k_y,\sigma}v^{\rm F}_{k_y,\bar{\sigma}}
/(v^{\rm F}_{k_y,\sigma}+v^{\rm F}_{k_y,\bar{\sigma}})$, $\gamma$ 
is the exponential of the Euler constant, $\Omega$ is the cutoff energy 
and $\Psi$ is the digamma function. If 
$k^{\rm F}_{k_y,\sigma}-k^{\rm F}_{k_y,\bar{\sigma}}+\alpha q_x=0$
is satisfied, $\tilde{K}_{k_y}(q_x)$ diverges logarithmically as $T$ goes
to zero. This logarithmic divergence survives over the $k_y$ summation in
eq. (10) only when Fermi surfaces for the up and down spins are 
``nested". This is not the case when the $k_y$ dependence of the Fermi
wave number is taken into account. The ``nesting" of the Fermi surface
becomes worse as $H$ increases, and as a result the critical magnetic field 
has a finite value.

In eq. (12), 
$k^{\rm F}_{k_y,\sigma}-k^{\rm F}_{k_y,\bar{\sigma}}$,
which is the wave number of the Cooper pair of electrons on the
left Fermi 
surface of down-spin and right Fermi surface of up-spin, 
gives the information of the ``nesting" condition as shown in Fig. 1(a).
In Fig. 1(b), we plot   
$a(k^{\rm F}_{k_y,\sigma}-k^{\rm F}_{k_y,\bar{\sigma}})$ for 
$t_b/t_a=0.1$, $t_{4b}/t_b=0$ and some values of higher harmonic terms
$t_{2b}$ and  $t_{3b}$ at $H=5$T.
We write the $x$-component 
of ``nesting vector" or the wave number of the Cooper pair ${\mib q}$ as
$q_x(s)=2s\mu_{\rm B}H/v_{\rm F}$, and plot $aq_x(s)$ in Fig. 1.
  
\begin{figure}
 \leavevmode
 \epsfxsize = 7.5cm
 \epsfbox{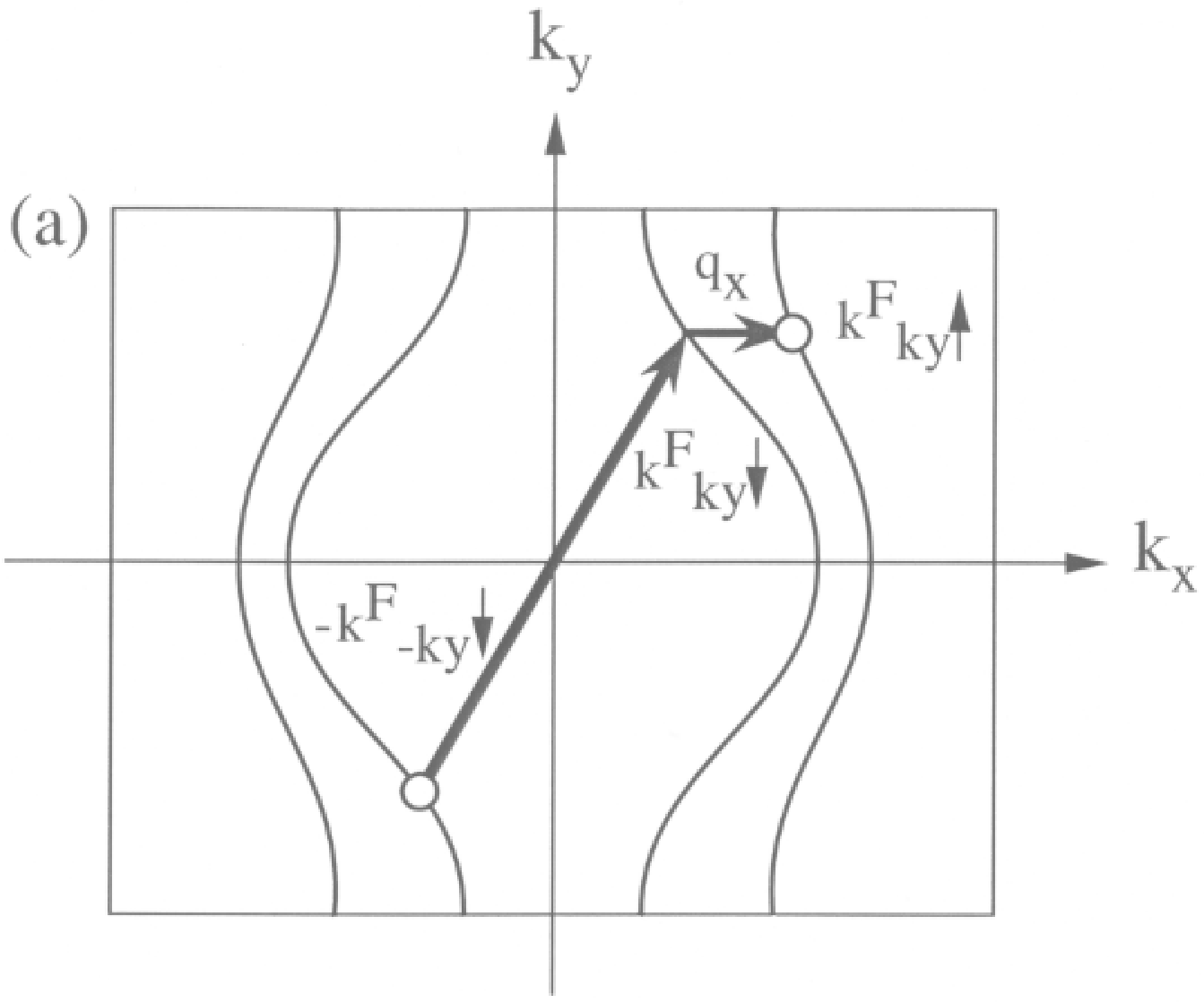}\\
 \epsfxsize = 7.5cm
 \epsfbox{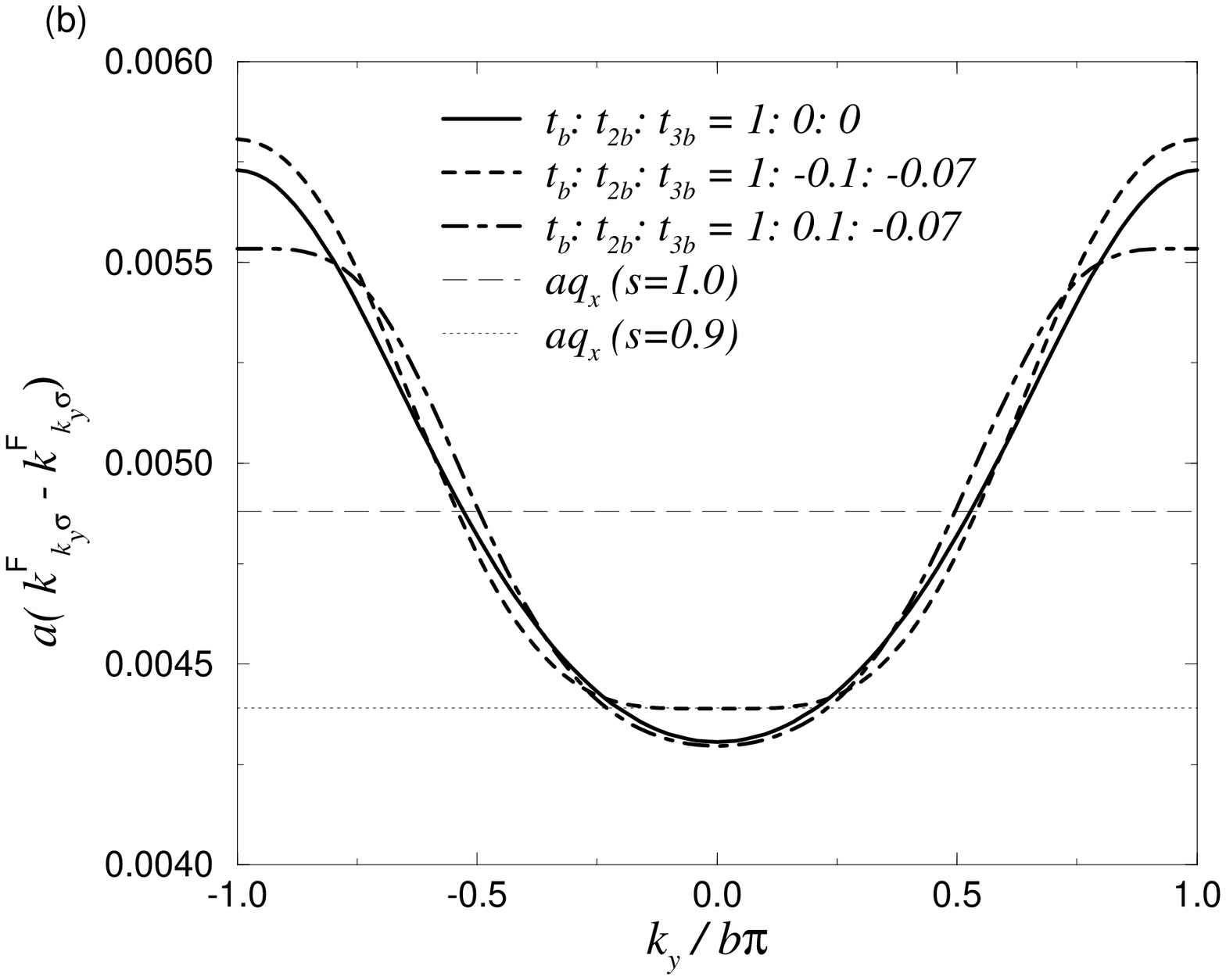}
\caption{(a) Schematic Q1D Fermi surface in $k_x$-$k_y$ plane in a 
magnetic field. 
(b) $a(k^{\rm F}_{k_y,\sigma}-k^{\rm F}
_{k_y,\bar{\sigma}})$ as a function
of $k_y/(b\pi)$. We take
parameters as $t_b/t_a = 0.1$, $t_{4b}/t_b=0$, quarter filled 
band and $H=5$Tesla. The long-dashed (dotted) line represents the 
$x$-component $aq_x(s)=2sa\mu_{\rm B}H/v_{\rm F}$ of the ``nesting" 
vector a${\mib q}$ in $s=1.0 (0.9)$.}
\label{fig1}
\end{figure}

In Fig. 2, we plot $T_{\rm c}(H)$ of the isotropic superconductivity 
obtained from  
eq. (9). In the following, we take parameters as $2t_a=1950$K, $t_b/t_a=0.1$,
$t_{4b}/t_b=0$, $T_{\rm c}(0)=1.35$K. The maximum
value of
$T_{\rm c}(H)$ for each $H$ is obtained by optimizing $q_x(s)$.

The lines with squares, solid circles and open circles in Fig. 2 are 
obtained for $t_c/t_a=0$, which corresponds to no orbital effect. 
The optimized $T_{\rm c}(H)$ in the absence of
higher harmonic terms is plotted by the line with solid circles. We find that 
the critical field at $T=0$,
$H^{\rm LOFF}_{\rm P}\sim 6.0$T is more than two times larger than 
the Pauli
paramagnetic limit $H_{\rm P}=1.84T_{\rm c}(0)\cong 2.5$T. This result is also
larger than that of the squares, where the nesting vector 
${\mib q}$ is fixed to be $2\mu_{\rm B}H/v_{\rm F},(s=1)$ as studied by
Lebed~\cite{rf:5}
($H^{\rm LOFF}_{\rm P}\cong 0.6\sqrt{t_a/t_b}H_{\rm P}\sim 4.7$T). 
By adding the higher harmonic terms $t_{2b}/t_b=-0.1$ and 
$t_{3b}/t_b=-0.07$, $H^{\rm LOFF}_{\rm P}$ becomes larger 
(the open circle line in Fig. 2, $H^{\rm LOFF}_{\rm P}\sim 9$T).
The enhancement of $T_c(H)$ and $H^{\rm LOFF}_{\rm P}$ due to the 
optimization of ${\mib q}$ and the higher harmonic terms ($t_{2b}<0$)
can be understood by the ``nesting" of the Fermi surface as follows. 
At low temperatures, only electrons with $v^{\rm F}_{k_y,\sigma}|k^{\rm
F}_{k_y,\sigma}-k^{\rm F}_{k_y,\bar{\sigma}}+\alpha q_x|<\delta\sim T$
can contribute to forming Cooper pairs. The
area on the Fermi surface, where $v^{\rm F}_{k_y,\sigma}|k^{\rm 
F}_{k_y,\sigma}-k^{\rm F}_{k_y,\bar{\sigma}}+\alpha q_x|<\delta$,
is proportional to $\sqrt{\delta}$ as $\delta\rightarrow0$ if down Fermi
surface touches the up Fermi surface by the translation of ${\mib q}$,
while it is proportional to $\delta$ if two Fermi surfaces cross by the
translation. Since $\sqrt{\delta}\gg \delta$ for $\delta\rightarrow0$,
$H^{\rm LOFF}_{\rm P}$ becomes largest when the 
Fermi surfaces touch by the translation, which is equivalent to the case
when 
$q_x(s)$ touches $k^{\rm F}_{k_y,\sigma}-k^{\rm F}_{k_y,\bar{\sigma}}$
in Fig. 1(b).
The enhancement of $H^{\rm LOFF}_{\rm P}$ in 2D by this mechanism has
been studied by Shimahara.~\cite{rf:25} These are two possibilities of
choosing optimal $q_x(s)$, i.e., $q_x(s)$ touches 
$k^{\rm F}_{k_y,\sigma}-k^{\rm F}_{k_y,\bar{\sigma}}$ at $k_y\approx 0$
or $k_y\approx \pm\pi/b$. It may depend on the curvature of the Fermi
surface and the Fermi velocity that which $q_x(s)$ gives the higher 
$H^{\rm LOFF}_{\rm P}$. We found that $H^{\rm LOFF}_{\rm P}$ is the 
largest when $q_x(s)$ touches at $k_y\approx 0$.
If we take the positive $t_{2b}$, the ``nesting"
becomes worse at $k_y\approx 0$, although it becomes better at 
$k_y\approx \pm\pi/b$ as shown by the dot-dashed line Fig. 1(b). In this
case $H^{\rm LOFF}_{\rm P}$ is not enhanced. We get 
$H^{\rm LOFF}_{\rm P}\approx 5.8$T for $t_{2b}/t_b=0.1$ and 
$t_{3b}/t_b=-0.07$. On the other hand $k^{\rm F}_{k_y,\sigma}-k^{\rm
F}_{k_y,\bar{\sigma}}$ becomes flatter at $k_y\approx 0$ when $t_{2b}$ is
negative as shown by the dashed line in Fig. 1(b), resulting in the
better ``nesting" and larger $H^{\rm LOFF}_{\rm P}$, as shown by the
line with open circles in Fig. 2.
 
Next, we take the effect of orbital motions into account.
Since the field dependence of the initial slope 
${\rm d}H^b_{\rm c2}/{\rm d}T|_{T_{\rm c}(0)}$ is approximately given by 
the parameter $t_ct_a$~\cite{rf:3,rf:5} in the week field limit,
we take parameters as $2t_a\sim 1950$K, $2t_c\sim 3.0$K, $a\sim 7\AA$,
$c\sim 13\AA$ and $T_{\rm c}(0)\sim 1.35$K in order to fit the initial slope
observed in (TMTSF)$_2$ClO$_4$~\cite{rf:8,rf:9}.
The transition temperature is plotted as the thick solid line in Fig. 2. This
curve seems to be consistent with the experiments in organic
superconductors (TMTSF)$_2$X by Lee {\it et al}~\cite{rf:8,rf:9}.

\begin{figure}
 \begin{center}
 \leavevmode
 \epsfxsize = 7.5cm
 \epsfbox{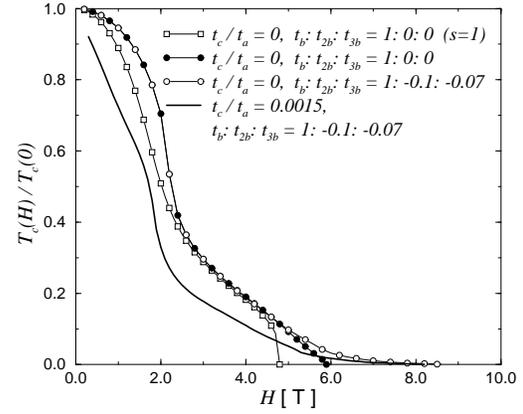}
\caption{Transition temperature $T_{\rm c}(H)$ of an isotropic pairing as a
function of the magnetic field in the case of $t_b/t_a = 0.1$,
$t_{4b}/t_b=0$,
$T_{\rm c}(0)=1.35$K for quarter filled electrons. The lines with 
squares and open circles
and thick solid line are obtained from the optimal $q_x(s)$ giving the 
maximum value of $T_{\rm c}(H)$. If $s$ is fixed in $1.0$, 
we get the line with solid circles.}
\label{fig2}
 \end{center}
\end{figure}
 
We also study $T_{\rm c}(H)$ of an anisotropic spin-singlet state.
As shown in our previous paper~\cite{rf:5}, the linearized gap 
equation for the anisotropic spin-singlet state is given by 
\begin{eqnarray}
& &\Delta(x)=2UT\sum_{k_y,k_z}\cos^2(ck_z)\sum_{\alpha,\omega_n}
\int_{|x-x'|>d} {\rm d}x'\Delta(x')\nonumber\\
& &\times G^{\alpha}_{\uparrow}(x,x',k_y,k_z,\omega_n)
G^{\bar{\alpha}}_{\uparrow}(x,x',-k_y,-k_z,-\omega_n).
\end{eqnarray}
where $U$ is the nearest-site attractive interaction along the $c$ axis. 
The energy gap is zero at the lines $|ck_z|=\pi/2$ in this model.
As in the isotropic case, eq. (13) is written as a 
matrix equation
\begin{equation}
\Delta^Q_{2l}=2U\sum_{l'}\sum_N\sum_{k_y}D^N_{l,l'}(k_y)
\tilde{K}_{k_y}(Q+NG)\Delta^Q_{2l'},
\end{equation}
where $\tilde{K}_{k_y}(q_x)$ is given in eq. (12) and 
the coefficients $D^N_{l,l'}(k_y)$ is defined by
\begin{equation}
D^N_{l,l'}(k_y)=\frac{1}{4}\left[M^1_{N,l,l'}(k_y)+
M^{-1}_{N,l,l'}(k_y)+2M^0_{N,l,l'}(k_y)\right],
\end{equation}
where $M^j_{N,l,l'}(k_y)$ is given by 
\begin{eqnarray}
M^j_{N,l,l'}(k_y)&=&\sum_nJ_{n+l}(\eta^{\alpha}_{k_y,\sigma})
J_{n+l'+j}(\eta^{\alpha}_{k_y,\sigma})\nonumber\\
&\times&J_{n-l+N}(\eta^{\alpha}_{k_y,\bar{\sigma}})
J_{n-l'+N-j}(\eta^{\alpha}_{k_y,\bar{\sigma}}).
\end{eqnarray} 

In Fig. 3, we plot $T_{\rm c}(H)$ of an anisotropic spin-singlet 
superconductivity. Since  
the initial slope of the anisotropic spin
singlet state is $\sqrt{2}$ times larger than that in the isotropic
state~\cite{rf:5}, we take parameters as
$2t_a\sim 2070$K and $2t_c\sim 4.0$K in order to fit the experimental
results. Other parameters are same as in the isotropic pairing case. 

\begin{figure}
 \begin{center}
 \leavevmode
 \epsfxsize = 7.5cm
 \epsfbox{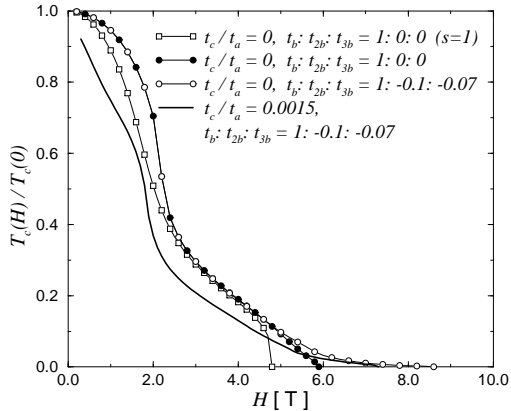}
\caption{
Transition temperature $T_{\rm c}(H)$ of the anisotropic pairing as a function
of the magnetic field.}

\label{fig3}
 \end{center}
\end{figure}

If $U=\lambda$ and $t_c=0$ ($\eta^{\alpha}_{k_y,\sigma}=0$), 
$T_{\rm c}(H)$ of an
anisotropic spin-singlet is same as an isotropic pairing since 
$S^N_{l,l'}(k_y)=D^N_{l,l'}(k_y)=\delta_{l0}\delta_{l'0}\delta_{N0}$. 
When $t_c\neq 0$, the behavior of 
$T_{\rm c}(H)$ for an anisotropic spin-singlet state is different from
that for the isotropic case, but the difference between them is small. 

In conclusion, we have calculated $T_{\rm c}(H)$ of the isotropic 
and anisotropic spin-singlet superconductivity in Q1D electrons.
Although the Zeeman splitting strongly 
suppresses the
superconductivity, the critical magnetic field is
enhanced by choosing the optimal ``nesting vector" and taking account of
the higher harmonic terms in the energy dispersion.
This result seems to be consistent with 
$T_{\rm c}(H)$ observed in the (TMTSF)$_2$X~\cite{rf:8,rf:9,rf:10},
which may show not only the possibility of spin-singlet pairing 
in this system but importance of higher harmonic 
terms.  

\section*{Acknowledgment}
One of the authors (K. K) was partially supported by Grant-in-Aid for JSPS 
Fellows from the Ministry of Education, Science, Sports and Culture.  
K. K was financially supported by the Research 
Fellowships of the Japan Society for the Promotion of Science for Young 
Scientists.

\end{document}